# First-principles study on the stability and electronic structure of monolayer GaSe with trigonal-antiprismatic structure

*Hirokazu Nitta[1], Takahiro Yonezawa[1], Antoine Fleurence[1], Yukiko Yamada-Takamura[1], and Taisuke Ozaki[2]*

[1]School of Materials Science, Japan Advanced Institute of Science and Technology (JAIST), 1-1 Asahidai, Nomi, Ishikawa 923-1292 Japan

[2]Institute for Solid State Physics, The University of Tokyo, 5-1-5 Kashiwanoha, Kashiwa, Chiba 277-8581, Japan

**Abstract**

The structural stability and electronic states of GaSe monolayer with trigonal-antiprismatic (AP) structure, which is a recently discovered new polymorph, were studied by first-principles calculations. The AP phase GaSe monolayer was found stable, and the differences in energy and lattice constant were small when compared to those calculated for a GaSe monolayer with conventional trigonal-prismatic (P) structure which was found to be the ground state. Moreover, it was revealed that the relative stability of P phase and AP phase GaSe monolayers reverses under tensile strain. These calculation results provide insight into the formation mechanism of AP phase GaSe monolayers in epitaxially-grown GaSe thin films.

**I. Introduction**

Two-dimensional materials exhibit many unique physical properties compared to bulk materials. In recent years, the study of layered metal-chalcogenides (LMCs) has been a topic of high interest because they exhibit a wide variety of properties depending on composition and number of layers[1,2]. The bonding between the atoms in a monolayer of LMCs is of covalent and/or ionic type, while the bonding between the layers is of the molecular, van der Waals type.

Gallium selenide (GaSe) is a LMC with a 2 eV band gap, and is known as a good nonlinear optical crystal owing to its non-centrosymmetric crystal structure[3]. A monolayer of GaSe is composed of covalently-bonded quadruple atomic layers in a Se-Ga-Ga-Se sequence, as illustrated in Fig. 1(a). This is for the trigonal prismatic GaSe structure. We name this GaSe crystal which adopts a conventional, wurtzite-like structure, "prismatic (P) phase", because Se atoms are coordinated in the form of a triangular prism with respect to the Ga dimer as shown in Fig. 1(a). Bulk GaSe with monolayers stacked vertically *via* van der Waals forces crystallizes in several polytypes with different stacking sequences: $\beta$-GaSe, $\varepsilon$-GaSe, $\gamma$-GaSe, and $\delta$-GaSe[3,4]. The most commonly found polytypes are ε- and γ-GaSe[5]. Recently, GaSe has been predicted to possess promising properties: ε-GaSe is expected to turn into a 3D topological insulator when a tensile strain of 3% or more is applied[6], while a hole-doped monolayer shows tunable ferromagnetism and half-metallicity[7]. Furthermore, a spin orbit coupling 10 times stronger than that for GaAs was observed experimentally for electron-doped ε-GaSe thin flakes with 10 and 25 nm thicknesses[8].

We have recently succeeded in growing epitaxial GaSe(0001) thin films on Ge(111) substrates using molecular beams of Ga and Se[9] through van der Waals epitaxy[10]. In a previous seminal paper, we reported the first experimental observation by high-angle annual dark field – scanning transmission electron microscope (HAADF-STEM) images[9] of single GaSe layer with a structure different from the one reported so far exists near the GaSe(0001)/Ge(111) interface. In this new structure, Se atoms are coordinated in a trigonal-antiprismatic way with respect to the Ga dimer as shown in Fig. 1(b). We name this GaSe crystal with this new structure, "antiprismatic (AP) phase". Monolayer AP phase GaSe has a centrosymmetric crystal structure in contrast to the non-centrosymmetric crystal structure of monolayer P phase GaSe.

The structural relationship between the P-phase and AP-phase GaSe monolayers is reminiscent of that between trigonal-prismatic and octahedral structures in transition-metal dichalcogenides (TMDCs) which are also LMCs. Depending on conditions such as alkali metal intercalation, both structures can be stabilized, which have very different properties[11–13]. For example, trigonal-prismatic $MoS_2$ is semiconducting, while octahedral $MoS_2$ is metallic[14,15]. On the other hand, for group- Ⅲ metal monochalcogenides such as GaSe, the variation in intralayer structure has hardly been discussed. So far, only one theoretical report presented the results of calculations on monolayer indium chalcogenides with antiprismatic structure named β-InX[16]

In this paper, we report the structural stability and electronic states of AP-phase GaSe monolayer obtained by first-principles calculations based on the density functional theory (DFT). Comparing the results of calculations carried out for P phase, we discuss the relative stability of the two phases, and the possible formation mechanism of the newly found AP-phase GaSe.

**II. Computational detail**

DFT calculations have been performed using the OpenMX code[17,18]. This code is based on norm-conserving pseudopotentials and optimized pseudoatomic basis functions[19]. The exchange-correlation functional was treated within generalized gradient approximation by Perdew, Burke and Ernzerhof (GGA-PBE)[20,21]. The regular mesh of 300 Ry in real space was used for numerical integrations. A (7×7×7) k-point mesh was used to discretize the first Brillouin zone in this study. The density of states (DOS) has been calculated with a tetrahedron method on a (12×12×12) k-point mesh. All atomic positions have been relaxed until the residual force on each atom has reached values of less than 0.0003 Hartree/Bohr. The vacuum space along the $z$-direction is taken to be more than 15 Å to avoid the spurious interaction.

**III. Results and Discussion**

**A. Structural stability**

The calculated total energies per chemical formula unit versus in-plane lattice constants of P-phase and AP-phase -GaSe monolayers are plotted in Fig.2. The lattice constants resulting in the lowest energies for P-phase and AP-phase GaSe monolayers were determined to be 3.81 Å and 3.82 Å, respectively. Note, that the experimentally obtained lattice constant of bulk ε-GaSe crystal is 3.74 Å[22]. With these stable lattice constants, the P-phase is more stable compared to the AP-phase. Note, however, that the energy difference between the two phases at the lattice constants of 3.81 Å is approximately 8 meV per formula unit. This energy difference is about the same as the cohesive energy difference (13 meV per unit cell), calculated for α–(P-phase) and β–(AP-phase) InX monolayers[16]. It is known that wurtzite (WZ) and zinc-blende (ZB) structures of GaN coexist in an epifilm grown by molecular beam epitaxy (MBE)[23]. The coordination of WZ and ZB structures are similar to the P-phase and AP-phase GaSe, respectively. In the case of GaN, the energy difference between WZ and ZB was reported to be in the range of 9 meV/f.u.[24] to 20 meV/f.u.[25]. The energy difference between the P- and AP-phases of GaSe is similar or less than that between two polymorphs of GaN. Since the energy difference is so small, the AP phase GaSe may have been formed in a non-equilibrium vapor growth process as in the case of GaN. This is consistent with the experimental observation of coexistence of the P-phase and AP-phase GaSe in thin films grown by MBE[9].

Furthermore, the energy difference between trigonal-prismatic and octahedral structures in TMDCs (e.g. MoS$_2$, MoTe$_2$, and WS$_2$) is about 0.5-0.9 eV/f.u.[26] which is about 100 times larger than that of GaSe or β–InX. It is assumed that the structural stability between the two structures is determined by the balance between the chalcogen-metal bond and the ion repulsion between the chalcogens[27,28]. Since the distance between the chalcogen atoms in the monolayers of III-VI layered materials are larger than that for TMDCs, the interaction between chalcogen atoms should be small. Therefore, the small calculated energy difference between the two structures of GaSe compared to those for TMDCs is consistent with the earlier studies[27,28].

The energy difference between the two phases decreases by increasing the lattice constants, and the relative stability reverses at the lattice constant of 3.96 Å. The result tells us, that, although the energy difference of the two phases is very small, the P-phase is more stable than the AP-phase at equilibrium lattice constants, which is consistent with the experimental observations[9]. On the other hand, the AP-phase tends to become more stable than the P-phase as the in-plane lattice constant increases. In other words, the calculation result suggests that the AP-phase could be stabilized by the in-plane tensile strain.

In our previous study, the AP-phase GaSe layer was observed at the first or the second layer from the Ge(111) substrate surface in HAADF-STEM images[9]. The optimized lattice parameter of in-plane Ge(111) surface calculated for bulk under the same condition was 4.09 Å, which is 7% larger than the optimized lattice constant of monolayer GaSe. At the GaSe(0001)/Ge(111) interface, most of the strain resulting from the difference between the lattice constant of Ge substrate and GaSe thin film is considered to be mitigated by the van der Waals interaction. However, our experimental observation and calculation results suggest that there could be a tensile environment at the vicinity of the film-substrate interface even in the van der Waals epitaxy growth.

**B. Band structure**

Electronic band structures for P-phase and AP-phase GaSe monolayers have been calculated for their optimized structures with lattice constants giving lowest energies. The resulting dispersion relations for the two phases, plotted along the high-symmetry directions of the two-dimensional hexagonal Brillouin zone (Γ-K-M-Γ), are shown in Fig.3(a) and (c). The zero energy is adjusted to the top of the valence band.

Both phases are indirect-gap semiconductors, primarily due to the valence-band maximum lying between the Γ and K points. The indirect-gaps of P- and AP-phases are 1.93 and 1.81 eV, respectively. In contrast to the band structures of bulk GaSe crystals, the valence-band maximum (VBM) has a local minimum at Γ point in monolayer GaSe which is sometime called "sombrero" dispersion[29].

The valence band of the AP-phase GaSe is similar to that of P-phase GaSe near the Γ point. However, some differences arise at the K point, where a doubly-degenerate band appears at the second and third highest valence bands and second lowest conduction band. In addition, the VBM at M point has an energy about 0.3 eV higher in the AP-phase than in the P-phase. Since both bands are mainly composed of Se $p_z$, the difference in band structure can be explained by the breaking of mirror symmetry.

The DOS of P-phase and AP-phase GaSe monolayers are shown in Fig.3(b) and (d), respectively. The DOS of AP-phase in valence band is similar to that of P-phase GaSe. Both results show a sharp van Hove singularity at VBM, similar to those which have been discussed in monolayer P-phase GaSe, GaS, and InSe, originating from the ring-shaped band extremum[29]. Ferromagnetism and half-metallicity are expected to emerge in monolayer P-phase GaSe with a large DOS at the Fermi level by hole doping[7]. Emergence of such properties can be also expected for monolayer AP-phase GaSe.

**IV. Conclusions**

Through first-principles calculations, we found that the AP-phase GaSe monolayer, which is a new polymorph, is metastable with a very small energy difference to the conventional P-phase GaSe, and this relative stability can be reversed under in-plane tensile strain. This result is consistent with the experimental observation that AP-phase GaSe monolayer was observed only near the Ge(111) substrate. The band structures of both phases were similar, but in the case of AP-phase GaSe, the bands with strong Se orbital character degenerates at the K point. The indirect band gap of AP-phase GaSe is about 0.1 eV smaller than that of P-phase GaSe.

This new phase GaSe may have been formed in previous growth experiments, especially in non-equilibrium processes. Actually, AP-phase-like GaSe can be seen in published STEM images, but not discussed at all[30]. The existence of AP-phase GaSe which has centrosymmetric crystal is expected to shed light in the understanding of various experimental results, for example, in the field of nonlinear optics.

**Acknowledgements**

This work was supported by Shibuya Science Culture and Sports Foundation. A part of this work was carried out using the facilities in JAIST, supported by Nanotechnology Platform Program (Molecule and Material Synthesis) of the Ministry of Education, Culture, Sports, Science and Technology (MEXT), Japan. H.N. and A.F. acknowledge support from the joint research program of the Institute for Solid State Physics, the University of Tokyo.

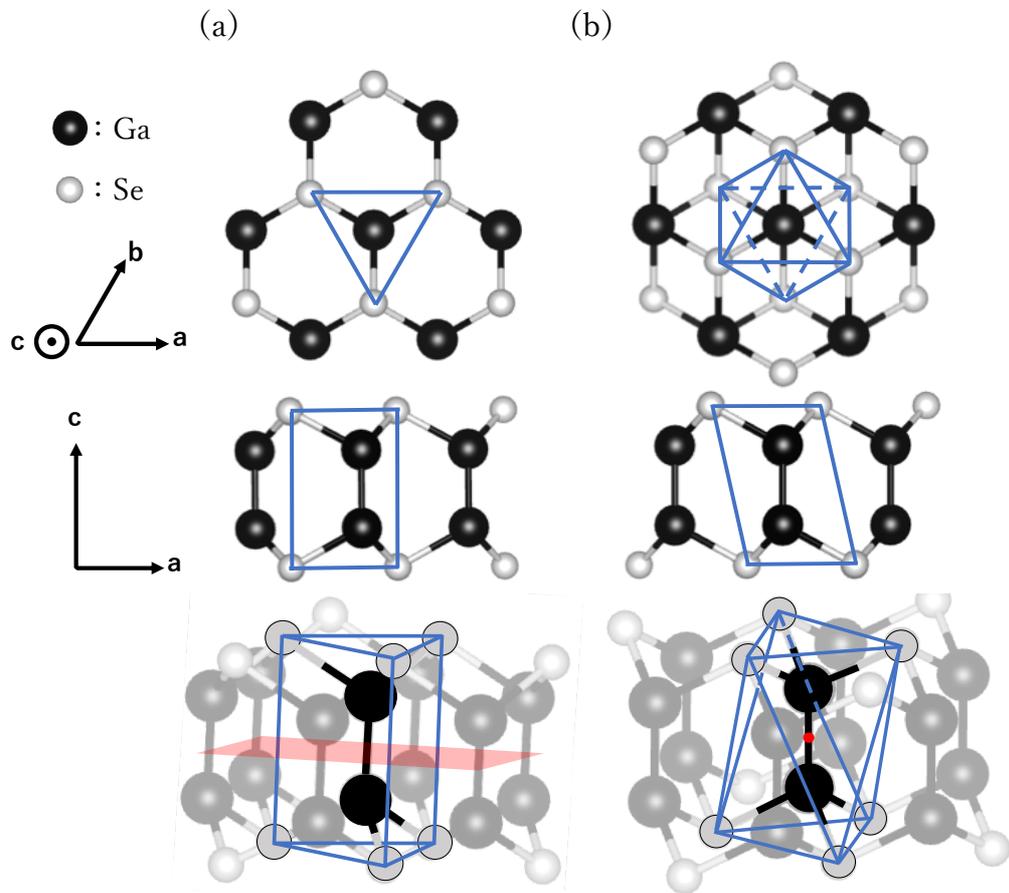

Fig.1 Top view (top), side view(middle) and perspective view (bottom) of crystal structures of (a) prismatic (P) -phase, and (b) antiprismatic (AP)-phase GaSe monolayers. Blue thin lines highlight (a) the triangular-prism and (b) –antiprism. P phase has mirror symmetry (mirror plane in red in the perspective view of (a)) and no inversion symmetry, while AP-phase has inversion symmetry (inversion center indicated as red point in the perspective view of (b)).

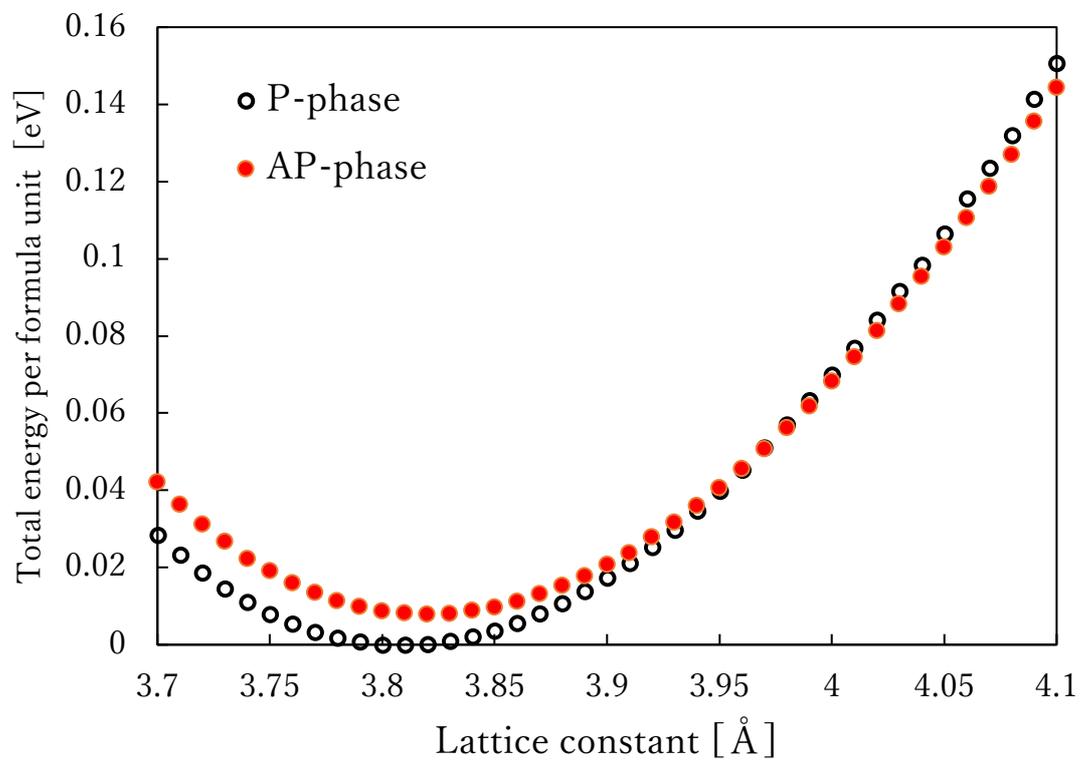

Fig.2 Energy versus lattice constant curves for P- and AP-phase monolayer GaSe. The lattice constants of the structurally-optimized P-phase and AP-phase GaSe monolayers are 3.81 Å and 3.82 Å, respectively.

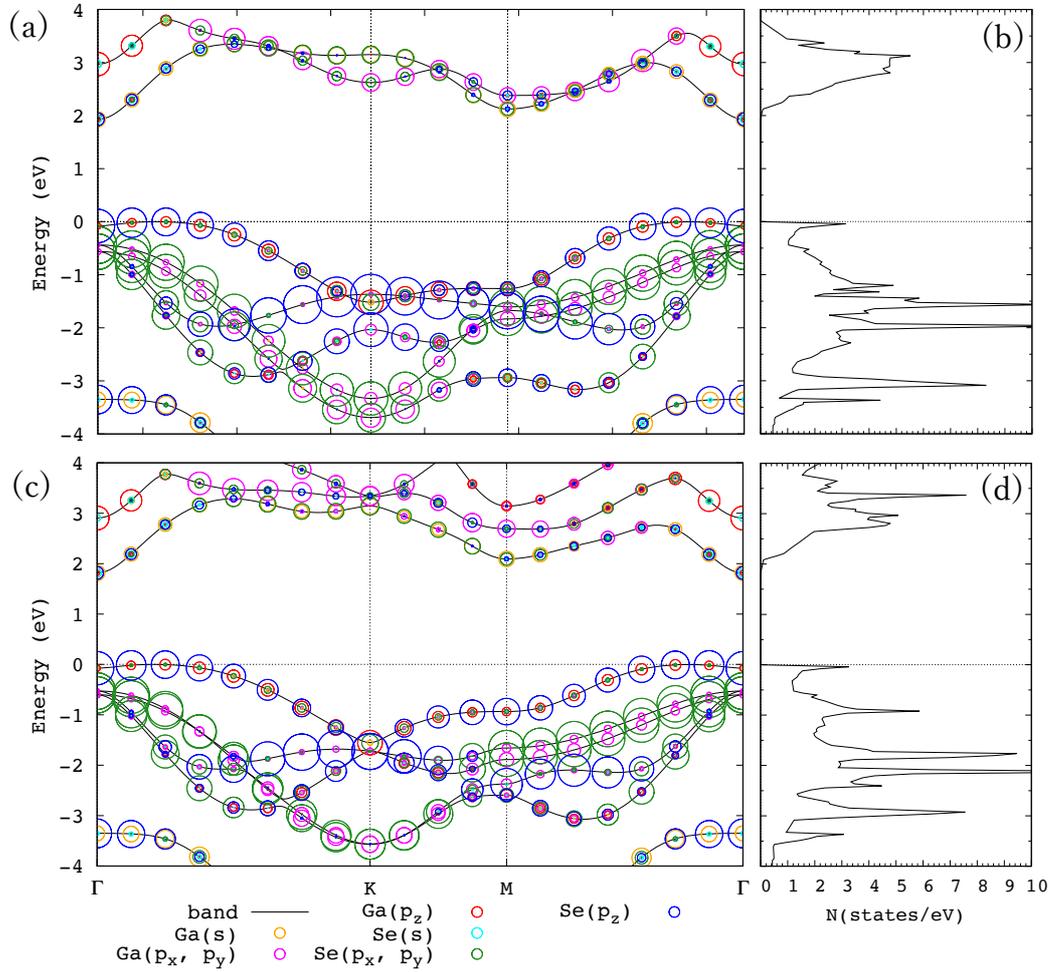

Fig.3 Band structures and density of states of (a), (b) P-phase GaSe, and (c), (d) AP-phase GaSe monolayers, respectively. Pseudoatomic orbital contribution is depicted on the band structures.